\documentclass[manuscript]{acmart}
\AtBeginDocument{%
  }

\acmDOI{XXXXXXX.XXXXXXX}

\usepackage{xspace}
\newcommand{\prompthis}{PrompTHis\xspace}
\newcommand{\prompthistwo}{PromptMap\xspace}
\newcommand{\demodel}{Design-Exploration model\xspace}

\newcommand{\todo}[1]{\leavevmode{\color{orange}{[Todo: #1]}}}
\newcommand{\yuhan}[1]{\leavevmode{\color{magenta}{[Yuhan: #1]}}}
\newcommand{\kai}[1]{\leavevmode{\color{blue}{[Kai: #1]}}}

\def \cleanversion{} %
\ifx\cleanversion\undefined
  \ifx\noremark\undefined
  \else
    \renewcommand{\remark}[1]{} %
    \renewcommand{\mylabel}[1]{} %
  \fi
\else
  \renewcommand{\todo}[1]{} %
  \renewcommand{\yuhan}[1]{} %
  \renewcommand{\kai}[1]{} %
\fi

\begin{document}

\title{\prompthistwo: Supporting Exploratory Text-to-Image Generation}

\author{Yuhan Guo}
\email{yuhan.guo@pku.edu.cn}
\orcid{0009-0004-3857-7486}
\affiliation{%
  \institution{Peking University}
  \city{Beijing}
  \country{China}
}

\author{Xingyou Liu}
\email{psxxl23@nottingham.ac.uk}
\orcid{}
\affiliation{%
  \institution{University of Nottingham}
  \city{Nottingham}
  \country{UK}
}

\author{Xiaoru Yuan}
\email{xiaoru.yuan@pku.edu.cn}
\affiliation{%
  \institution{Peking University}
  \city{Beijing}
  \country{China}
}

\author{Kai Xu}
\email{kai.xu@nottingham.ac.uk}
\orcid{0000-0003-2242-5440}
\affiliation{%
  \institution{University of Nottingham}
  \city{Nottingham}
  \country{UK}
}

\renewcommand{\shortauthors}{Guo et al.}

\begin{abstract}
Text-to-image generative models can be tremendously valuable in supporting creative tasks by providing inspirations and enabling quick exploration of different design ideas. However, one common challenge is that users may still not be able to find anything useful after many hours and hundreds of images.
Without effective help, users can easily get lost in the vast design space, forgetting what has been tried and what has not.
In this work, we first propose the \textit{\demodel} to formalize the exploration process. 
Based on this model, we create an interactive visualization system, \textit{\prompthistwo}, to support exploratory text-to-image generation. Our system provides a new visual representation that better matches the non-linear nature of such processes, making them easier to understand and follow. It utilizes novel visual representations and intuitive interactions to help users structure the many possibilities that they can explore. %
We evaluated the system through in-depth interviews with users.
\end{abstract}



\begin{teaserfigure}
  \includegraphics[width=\textwidth]{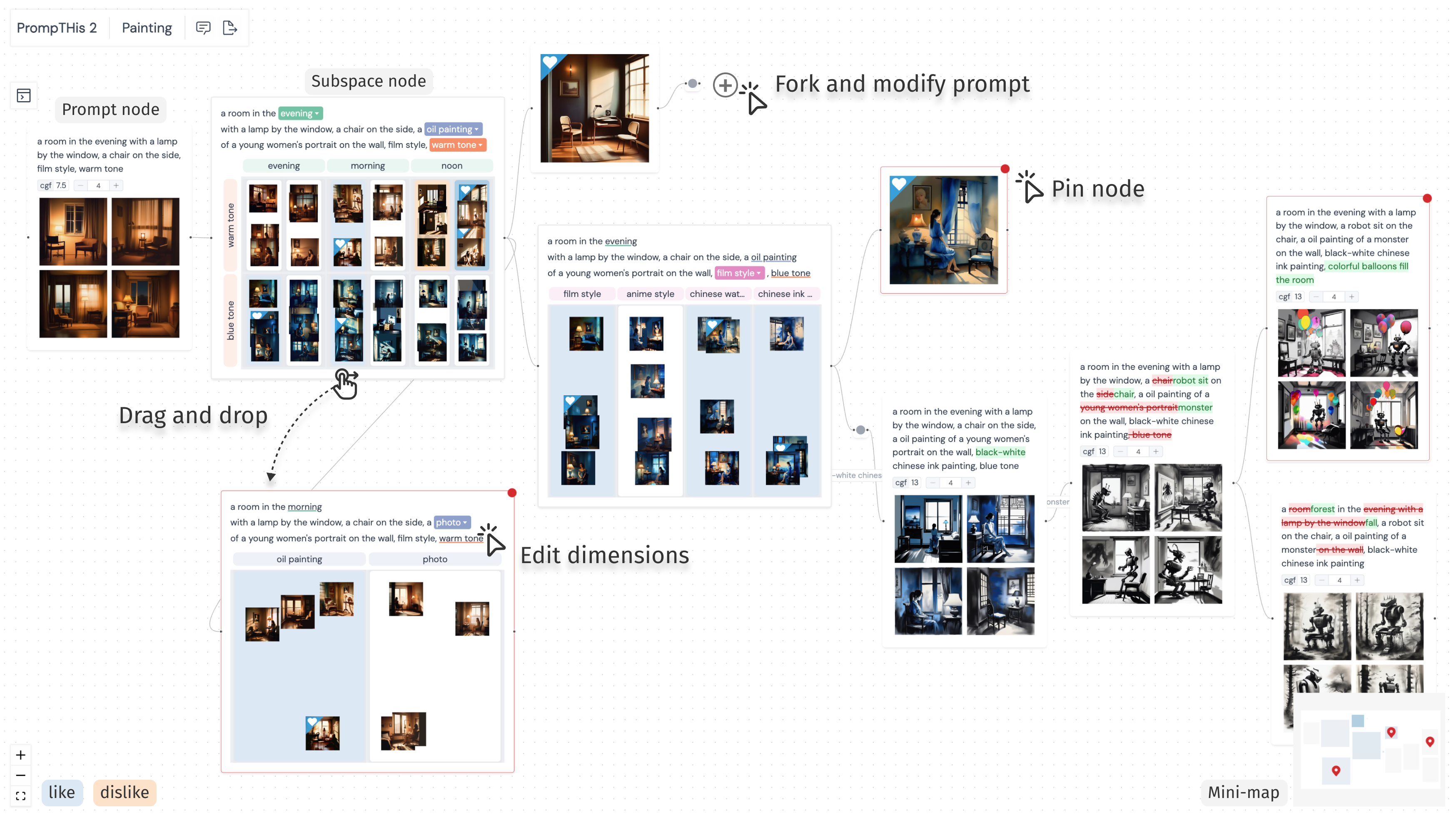}
  \caption{\prompthistwo captures the prompting process using a tree-like diagram, where the linkages represent that the child prompt is modified on the basis of the parent. The node can be either a prompt or a \textit{subspace}. The \textit{subspace} node supports users to explore different values of certain aspects of the prompt (\textit{dimensions}), such as subjects and styles, which is visualized as nested grids.}
  \label{fig:teaser}
\end{teaserfigure}

\maketitle

\section{Introduction}

Generative AI has been transforming the way that people explore and create. Using text-to-image models, users can generate a variety of images through textual prompts. This significantly reduces the cost for creators to test out their ideas and accelerates the iterative exploration. The occasionally unexpected yet inspiring outcomes also illuminates ``lateral paths''~\cite{liapis2016can} in the design space, thus expanding the breadth of the exploration and fostering human creativity.

Despite the enormous potential, the use of generative models can lead to increased cognitive and metacognitive demands on users~\cite{tankelevitch2024metacognitive}. As the model does not always understand the user's intent, a lot of generated results are not satisfactory and the users need to go through many trials and errors before they can obtain something useful. The lack of transparency and interpretability of generative models imposes a high cognitive demand to make sense of how these models work. Users spend much time refining specific prompts to convey their intentions, easily getting stuck or making repetitive efforts. At the same time, when the model produces interesting and inspiring results, users often struggle to manage multiple possibilities simultaneously, focusing on a few options and leaving a vast space of alternatives unexplored. As a result, creators often get lost in the creative process and engage in near-random exploration~\cite{guo2024prompthis}. Such an issue is also known as the \textit{disorientation} problem~\cite{ayers1995using,nguyen2016sensemap}.

Ever since the emergence of generative models, research efforts have been devoted to reducing users' cost in conveying their intentions to the models. The approaches include explaining the detailed generation process inside the model~\cite{lee2023diffusion}, helping users interpret the mapping from their prompts to images~\cite{guo2024prompthis}, recommending relevant prompts and images~\cite{feng2023promptmagician}, automatically refining user's prompts~\cite{wang2023reprompt, brade2023promptify}, and translating user inputs into specifications that explicitly guide the models~\cite{zeng2024intenttuner}. These methods improve the efficiency in getting the desired image but the disorganization issue remains unsolved when it comes to open-ended exploration. Also, it is still challenging for generative AI systems to fully understand users' intention or their styles and preferences without effective access to the context of user's exploration process.

An alternative approach is to actively assist users in structuring the exploration by recommending the design space and possible candidates to users. A common solution is using a schema to formalize the design space so that computational models can be applied to automate the process~\cite{suh2024luminate, almeda2024prompting}. The design space can be represented using dimensions, the orthogonal aspects or variables associated with the task. For example, in text-to-image generation, the dimensions can be the subject, medium, style, etc. As part of the exploratory process, new images are generated by changing the values for these dimensions. This helps structure the exploration and also encourages divergent thinking.
However, the actual design space can be very large, as the number of possible prompts increases exponentially with the number of dimensions. This makes it impossible to explore the entire design space and there is an urgent need for better support of such exploratory creation.
The existing attempts usually deploy a linear list to record such process, which does not match and then correctly reflect the characteristics of such exploration. Also, they are usually not designed to handle the scalability issue discussed earlier, leaving it to the users to manually manage this complex and often messy process.

This work aims to provide better support for users to manage and structure their creative exploration process.
It also gives them complete control of the exploration process, without introducing potential bias that may be embedded in the so-called ``popular'' prompts or introduced by the system recommendations.
We conducted interviews with generative AI users to understand how they use these models and what the challenges and needs are.
To meet these requirements, we first propose a \demodel to describe the exploration process. Inspired by the data-frame theory for the sensemaking process~\cite{klein2007dataframe}, the model represents the exploration process as iterative construction and elaboration of ``subspaces‘', which are a subset of the entire design space.
Based on this representation, we developed a visualization system to support such process. It captures the history of the exploration process, allowing users to systematically explore and refine promising subspaces.

Our system uses a node-link diagram to visualize the thinking process. To support flexible exploration, the node can either be a prompt or a structured subspace. To visualize the multi-dimensional subspace, we use the dimensional stacking~\cite{ward1994xmdvtool} method to represent the subspace node as nested grids. The system provides rich interactions for users to engage in creative exploration, such as forking nodes, editing dimensions, intuitive interactions with grid cells and images, and curating the results. We conducted in-depth interviews with users from different backgrounds to evaluate the system and understand how the design could assist in the creative exploration. The participants liked the node-link representation of the thinking process and the convenience of the new way to create and explore different dimensions, which leads to a more effective and organized creative experience.

To summarize, our contributions include:
\begin{itemize}
    \item The \demodel, which provides a formal representation of the exploration process.
    \item Based on this model, an interactive visualization system that supports the open nature of the creative process and addresses the scalability issue of vast design space through more structured exploration.
    \item In-depth interviews with users demonstrating how our system supports exploration.
\end{itemize}

\section{Related Work}

This section begins with some theoretical work on the creative process and how computers facilitate this process. The next part focuses on the related literature in the field of text-to-image generation. Finally, we review the recent work on developing interactive user interfaces for generative models to support exploration.

\subsection{Creative Process and Creativity Support Tools}
\label{sec:related-work-theory}

Creativity is often considered essential in a diverse range of domains such as scientific research, artistic design, among others. Tremendous amount of research has been done to study the mechanisms and factors of creativity~\cite{kaufman2019handbook}. In practice, creative work often emerges from the exploration of variables and alternatives~\cite{gero1990design, stebbins2001exploration}.
Guilford defined such ability to generate multiple answers as \textit{divergent thinking}~\cite{guilford1968creativity}, in contrast to \textit{convergent thinking}, that is, to arrive at one answer according to the task constraints.
Based on this definition, the creative process is considered as the iterative interplay between divergence and convergence~\cite{designcouncil2024framework, roberts2015sketching}.

Creativity support tools can aid the creative process by suggesting alternatives that empowers divergent thinking~\cite{shneiderman2007creativity}. The term ``mixed-initiative co-creativity'' was introduced to emphasize the potential of computers to actively engage in creative activities by rapidly exploring a vast possibility space under the guidance of human's creative thinking~\cite{yannakakis2014mixed, deterding2017mixed, koch2020imagesense}. Furthermore, the recent advancements in deep generative models have enabled the computer to \textit{generate} diverse options even when there is no well-defined design space, therefore expanding the support for co-creativity to a wider variety of media such as images and videos~\cite{forbes2020creative}.

Though considered essential to the creative process, exploration does not guarantee creativity. According to Gero's model of design~\cite{gero1990design}, creative design involves the use of new variables that extend the design space. Similarly, the value of computer suggestions to creativity lies in the random stimuli they provide that could re-frame the design space~\cite{liapis2016can}. From this perspective, a fundamental part of the creative process is \textit{sensemaking}~\cite{weick1995sensemaking, davis2017creative}, i.e., the creator gets familiar with and transforms the design space through understanding the explored alternatives.

\subsection{Text-to-Image Generation and Prompt Engineering}
\label{sec:related-work-t2i}

Text-to-image generative models take natural language prompts as input and generates images that match the prompt description. A lot of text-to-image models, such as VQGAN-CLIP~\cite{crowson2022VQGAN} and latent diffusion~\cite{rombach2022high}, are based on CLIP~\cite{radford2021learning}, which aligns natural languages and images in vector-based representations. As these models can quickly generate images according to user description, many artists use them to test out artistic ideas and seek inspirations for creativity.

Despite the enormous potential, it could be challenging to compose prompts that can effectively convey user intentions to the model. The artists often need to go through trial and error to refine the prompts. Researchers and community users have been investigating strategies to search for effective prompts that can guide generative models to produce satisfactory results, which is known as \textit{prompt engineering}~\cite{liu2022design}. Part of these works present empirical guidelines by analyzing user prompts or experimental data~\cite{liu2022design, oppenlaender2023taxonomy, mahdavi2024ai}. The guidelines provide macro-level knowledge about generative models such as the template to use, but do not capture micro-level knowledge such as how specific wording works. Some other works support user to conduct ad-hoc experiments based on their own tasks and visually observe the results~\cite{mishra2023promptaid, strobelt2022interactive}. We adopt a similar approach in this work, i.e., allowing users to easily configure variables in the prompt and compare the output images.

As it still requires expert knowledge to come up with different variables for testing, systems that target non-expert users may directly recommend prompts and images to users. For example, RePrompt~\cite{wang2023reprompt} is a model customized for emotional expression, which automatically refines users' prompts with clear subjects and emotional descriptions. PromptMagician~\cite{feng2023promptmagician} retrieves similar prompt-images pairs based on user input and help users find interesting keywords. LLMs are also used for refining prompts~\cite{brade2023promptify, wang2024promptcharm}. However, for artists with distinct intentions and personal preferences, those dataset-based or similarity-based recommendations may not be a good fit as the recommended results might not match the artistic intentions.

Prompt engineering is not the only way to control the generation. Some works provide additional support such as inpainting generation~\cite{chung2023promptpaint, wang2024promptcharm}. The users can select specific areas in the image that they want to remove or replace. These methods are effective when the users have concrete demands but are less helpful for an exploratory setting, so we didn't include them in our prototype system.

\subsection{Creative Process Support for Generative Models}

While Section~\ref{sec:related-work-t2i} covers systems for improving specific prompts, this section presents an orthogonal perspective: user interfaces for generative models that support the entire exploratory workflow, which is also the focus of our work.

Generative models such as LLMs and text-to-image models can produce large amounts of information very quickly. This makes it difficult to manage, especially for complex tasks that require working in a nonlinear manner~\cite{suh2023sensecape, zhou2024understanding}. An example interface tailored to the nonlinear workflow is Dramatron~\cite{mirowski2023co}, a Colab-based system designed for screenwriters to organize the generation in a hierarchical structure, but the approach is not directly applicable to our use case. 

As mentioned in the Introduction section, a common challenge faced by the users when engaging in exploratory tasks is \textit{disorientation}. 
A basic approach to support users to orient themselves in the creative journey is providing rich history-keeping, which is one of the principles for designing creativity support tools~\cite{shneiderman2007creativity, weisz2024design}. Besides presenting the history versions as a linear list, \prompthis~\cite{guo2024prompthis} embeds the prompt-image pairs into a two-dimensional space to show the thematic distribution of the generated content. However, the projection-based representation in \prompthis does not reflect the detailed evolution of the user's thinking process.
Recently, canvas-based interfaces have been popular due to their support for flexible information management and spatial thinking.
For example, Sensecape~\cite{suh2023sensecape}, which is designed for information foraging and sensemaking with large language models (LLMs), supports users to organize the acquired information on multi-level canvases.
However, different tasks may require different representations and interactions, e.g., multi-level canvases may not be suitable for text-to-image generation as the generated content does not have a natural hierarchical structure.
A representation that can capture the creative process is needed to manage the exploration.

To encourage divergent thinking and reduce fixation, Luminate~\cite{suh2024luminate} emphasizes the strategy of \textit{structured exploration}, i.e., the model should allow dimensional exploration of the design space instead of responding with a single data point. The ``dimension'' idea can be used in text-to-image generation, e.g., the subject terms or style terms in the prompt can be treated as dimensions. However, as the prompt for text-to-image generation might include a lot of details that can be possible dimensions, the design space could be very large and it is impossible to explore the entire space. Instead of pre-defining the design space, DreamSheets~\cite{almeda2024prompting} provides a more flexible manner, using a spreadsheet for easier input of the dimensions.
Our work differs from the existing work in that, beyond structuring a single attempt, we aim to support the exploration process that involves multiple evolving attempts. Also, we believe that an ideal interface for exploratory tasks should both provide nonlinear history-keeping and support structured exploration, so it is important to understand how the thinking process evolves with the exploration. %

\section{Requirements}
\label{sec:design-rationale}

In this section, we describe the user practices and their requirements that motivates this work on representing the exploration process.

\subsection{User observation and interview}

We interviewed six users to observe how they use generative models and the challenges they face.
Four of the participants were professional artists. The other two were a designer and an experienced amateur user. All participants frequently use generative models to explore creative ideas.
The participants were provided with a web interface 
that includes an input box for users to specify the prompt and presents the prompt history as a scrollable list. The whole session took around 40 minutes, including a 25-min open-ended exploration session, where the participants chose a task from their usual workflow and demonstrated their workflow through the provided interface. This is followed by a 15-min semi-structured interview to understand their experiences.
The interviews were recorded and thematic analysis
was applied to the interview transcript. The results were synthesized to create the usage scenario and requirements below.

\subsection{Usage Scenario}
\label{sec:usage-scenario}

The users typically use the generative models for artistic ideation, such as visualizing and developing ideas for a film. They usually start with an ambiguous goal (e.g., an abstract concept) and gradually refine it through exploration. In the exploratory scenario, users use text-to-image models to test their ideas and seek for inspirations that may emerge from unexpected results. This process is different from using generative models for production, where users already have a clear target image in mind before the generation starts.

The main takeaway from the observed workflows is that the users \textbf{frequently interact with the exploration history}. This is because the exploration process involves a lot of trials and errors, as the generative models can produce a lot of images but many of which may be unsatisfactory so users frequently need to revert to earlier attempts. Besides locating the ``checkpoint'' for exploration, the users also review the previous attempts to discern the mechanism of the underlying model and reflect on potential alternative ideas. However, the web interface the participants used, which is similar to many popular UIs (such as that of Midjourney~\cite{midjourney}), presents the history as a linear list of prompts and generated images. As the number of iterations increases, the interaction with the prompt history becomes increasingly laborious and the artists find it difficult to find the desired attempt and manage the exploration. As a result, we decided to foucs our efforts on this.

\subsection{User Requirements}

Below we describe the specific user requirements regarding their interaction with the prompt history.

\textbf{R1. Reflect the evolution of user thinking.}
One of the main issues that slows down the creative workflow is that the users can easily lose track of the previous exploration attempts. The users might reach a dead end after a long series of attempts and have difficulty finding the way back and identifying appropriate points to branch out for new explorations. At other times, they might work on a few different ideas at the same time, switching among them. In this case, users can easily forget what has been explored or what is unfinished and end up repeating the attempts or overlooking otherwise promising leads. While keeping the prompt history as a list could help to some extent, it does not reflect the evolution of user thinking, which is nonlinear (e.g., with branches) by its nature. We see a need for a more effective approach that better matches the characteristics of the creative thinking process and supports more efficient review of the exploration history.

\textbf{R2. Help users compare and analyze the effects of prompt changes.}
The model does not always understand the user's intention in the prompt and the users rely on trial and error to adapt their prompts. The users would like to know how the underlying model works by experimenting with different wording and comparing the outputs. However, it is not easy to locate, organize, and analyze the attempts that they want to compare, as these can be located far apart, especially when multiple parts in the prompt are changed.

\textbf{R3. Support more efficient and structured exploration.}
The exploration may appear random, with no clear intention connecting consecutive steps. However, base on our observation, the artists do follow some prompting strategies such as they would replace a couple of keywords or key phrases in the prompt with different wordings. However, the existing tools do not capture and visualize such behavior patterns in a way that is easy to reason with. Another related challenge is that when several parts of the prompt are being modified, there are many combinations depending on the number of changing parts and possible values for each. It is difficult to manually enumerate all possible combinations, and equally difficult to track what has or has not been unexplored. The artists need a more efficient way to manage such exploration strategy.

\section{\demodel}
\label{sec:model}

Although exploration is common in creative practice and scientific investigation, the meaning of the term and the dynamics of the exploration process are not yet clearly understood.
In this section, we introduce the \textit{\demodel}, which provides a formal representation of the exploration process.
This model is a synthesis of existing theories on exploration-related activities such as design and sensemaking.

\subsection{What is Exploration}

Stebbins summarized the different definitions of \textit{explore} in dictionaries in his paper discussing exploratory research in social science~\cite{stebbins2001exploration}. According to Stebbins, in the most general sense, to explore means to study, examine, analyze, or investigate something. A definition that better fits the practice of artistic or creative design is ``to become familiar with something by testing it or experimenting with it''. Stebbins termed this \textit{innovative exploration}, as opposed to \textit{exploration for (scientific) discovery}, i.e., ``to travel over or through a particular space for the purposes of discovery and adventure''. The former stops when the desired result is achieved, while the latter is performed more systematically and aims for comprehensive discoveries.
Despite the distinction highlighted by Stebbins, in practice, exploration is often a combination of both processes. For example, a designer may travel through a certain design space and experiment with different alternatives to discover new ideas.
Based on the definitions, we identified three activities involved in the exploration process.

\begin{itemize}
\item \textbf{Experimentation} refers to testing or investigating a certain target, such as running the model with a prompt and examining the generated image.
\item \textbf{Navigation} refers to the travel over a certain spatial or conceptual space, such as traversing the space of possible prompts or creative ideas. It is often accompanied with \textit{experimentation}, i.e., testing the things encountered in the space.
\item \textbf{Interpretation} refers to the comprehension of experimentation results, such as the understanding of which ideas work or which prompts are effective.
\end{itemize}

\subsection{Existing Representations}
\label{sec:model:existing}

To investigate how the exploration process can be formally represented, we reviewed relevant representations of processes, including design and sensemaking, in which exploration plays an important role.

\textbf{Design.} A widely adopted framework for design or innovation is the Double Diamond model~\cite{designcouncil2024framework}. This model encourages creative practitioners to explore multiple answers (\textit{divergent thinking}~\cite{guilford1968creativity}) before taking focused actions (\textit{convergent thinking}) through iterative cycles of divergence and convergence.
One approach to think divergently is to navigate the \textit{Design Space}, i.e., the space of all possible designs.
In practice, a common representation of the design space is the Cartesian space, which is described by a set of orthogonal dimensions~\cite{suh2024luminate}, such as the style of the image.
As the design space is usually ill-structured and obscure to creators~\cite{gero1990design}, the Cartesian space is often an approximation of the design space. The actual design or exploration process usually consists of both unstructured and structured attempts, and the approximating structures are updated as the user's understanding of the design space evolves.

\textbf{Sensemaking.}
Sensemaking is the process that people ``structure the unknown''~\cite{ancona2012sensemaking}; consisting of information foraging and interpretation to construct meaning, the sensemaking process is often entangled with exploration. A fundamental model that describes the sensemaking process is the data-frame theory~\cite{klein2007dataframe}, which characterizes the sensemaking process as the dynamic interplay between data and cognitive \textit{frames} (the explanatory structure that accounts for the data and guides the search for additional data). Centered around the \textit{frame}, sensemaking activities include constructing a frame, elaborating a frame, questioning a frame, re-framing, among others.
While the exploration process and sensemaking process share similar activities (e.g., forage/navigate information and interpret/schematize knowledge) and are often intertwined, the focuses of the two process are different. The goal of sensemaking is to obtain a coherent structured account of the data (the \textit{frame}). In the exploration process, however, the \textit{interpretation} primarily serves as navigation aids, i.e., to figure out the next step, and the focus is the movement (\textit{navigation}) across space.

\begin{figure}[ht]
    \centering
    \includegraphics[width=\linewidth]{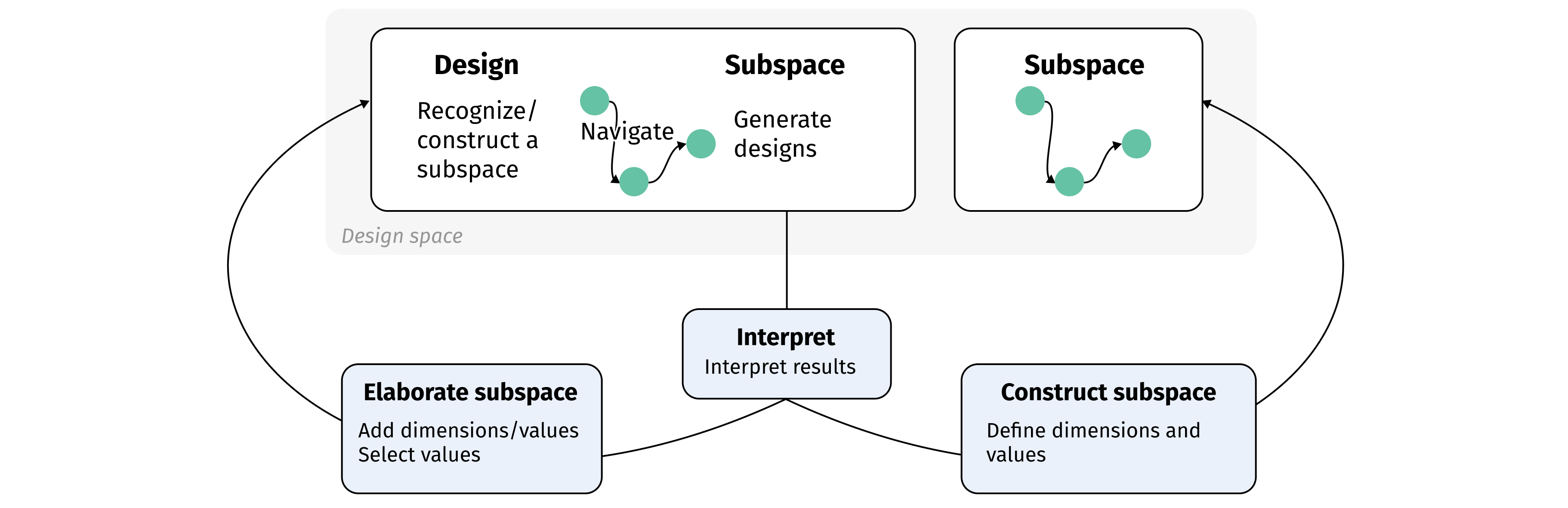}
    \caption{
        Design-exploration model (adapted from the data-frame model~\cite{klein2007dataframe}).
    }
    \label{fig:creative-activities}
\end{figure}

\subsection{Synthesized Model}

Certain commonalities exist across the design, sensemaking, and exploration processes. In the creative process, the candidate design can be seen as \textit{data}, and the structured representation of the \textit{design space} can be seen as a \textit{frame} that guides the navigation of the possible designs.
Such a \textit{frame} is referred to as \textbf{subspace} in our \demodel. Subspace denotes the structured subset of the design space that the user travels over.
As we have introduced in Section~\ref{sec:model:existing}, people often don't have a comprehensive mental model of the design space. Instead, the exploration usually focuses on a small subset of the design space. It is the specific subarea that people can examine and comprehend. In our model, the term \textit{subspace} implies that the space must be well-defined so that users can carry out systematic exploration over it.
Specifically, in the context of text-to-image creation, we define the subspace as a structured space of candidate prompts represented as a Cartesian space constructed by a set of dimensions, e.g., subject (the values include person, animal, character, etc.) or medium (the values include photo, painting, illustration, etc.).

The interplay between data and frame characterized by the data-frame model also offers an analogy for describing the dynamics of the exploration through interaction between designs and subspaces, that is, how the \textit{design space} is revealed and expanded during the exploration process. As shown in Fig.~\ref{fig:creative-activities}, the exploration process can be described as the navigation of the design space through dynamic construction and elaboration of the subspaces. The exploration either starts with the \textbf{initial construction of the subspace} that generates the designs, or the creation of the initial prompts, from which the creator may abstract a structure and \textbf{recognize the subspace}.
After examining and interpreting the generated designs, creators may choose to continue with ``within-subspace navigation'' or ``between-subspace navigation''. The former refers to the refinement of the idea through \textbf{elaboration of the subspace}, i.e., refining the schema by adding dimensions and values, or selecting desired values and filtering irrelevant ones. In other circumstances, such as when the results are unsatisfactory or a new idea comes up, creators can perform between-subspace navigation, taking on a different path by \textbf{constructing a new subspace}.
The major difference between our \demodel and the data-frame model for sensemaking is that, in sensemaking the frames are competing and selective, whereas subspaces represent different areas in the design space and exploration seeks for the extension of existing subspaces and construction of new subspaces. Therefore, activities such as reframing described in the data-frame theory are omitted in our model.

\section{\prompthistwo}

This section presents \prompthistwo, an interactive visual interface that applies the \demodel (Section~\ref{sec:model}) to address the user's requirements (Section~\ref{sec:design-rationale}).

\subsection{Design Goals}
\label{sec:design-goal}

The \textbf{usage scenario} that \prompthistwo targets is exploratory creation, i.e., the user starts with an ambiguous goal and gradually refines it through exploration. The \textbf{target user} of the system is users who use text-to-image models for ideation. These may include artists and designers, and also amateurs involved in similar creative processes but not as a profession. We assume the users are familiar with generative models and prompt engineering.
Based on the user requirements and relevant literature, we established the following design goals.

\begin{itemize}

\item \textbf{G1. Visualize the exploration history.}
The \demodel characterizes the exploration process as generating designs, constructing subspaces, and elaborating subspaces. We aim to visualize the history of such a process, i.e., visualizing the user's navigation paths within and between the subspaces, which reflects the evolution of their creative thinking (\textbf{R1}).

\item \textbf{G2. Visualize the multi-dimensional subspace.}
One of the main roles of the dimensional structure of the subspace is to allow users to analyze the influences of prompt changes (\textbf{R2}) by comparing different values of the dimensions. We aim to support the comparison by visualizing the structure of the subspace. The scalability issue should be addressed when there are multiple dimensions.

\item \textbf{G3. Provide intuitive interactions for managing the exploration.} In order to support users to carry out their exploration efficiently (\textbf{R3}), the system should provide fluid interactions for users to construct and elaborate the subspaces.

\item \textbf{G4. Integrate user curation with the visualization.}
The user's feedback is an important aspect of the exploratory process (\textbf{R1}). The information such as which images the users like or dislike is crucial to understanding the attempts that users made and should be reflected in the visualization.
Additionally, the feedback collected would allow the system to learn user's preferences and make recommendations in the future.

\end{itemize}

\subsection{Visual Encoding and Interactions}

\prompthistwo uses a node-link diagram (Fig.~\ref{fig:teaser}), where the nodes represent the subspaces and the links are the transition between the subspaces, to visualize the exploration history (\textbf{G1}). We designed a recursive grid representation to visualize the multi-dimensional subspaces and support the comparison of the dimension changes and the resulting outputs (\textbf{G2}). Intuitive interactions are introduced to support efficient and structured exploration (\textbf{G3}). Finally, user curation is also supported and encoded (\textbf{G4}). We will introduce these in detail in the following sections.

\subsubsection{Exploration Map}
\label{sec:exploration-map}

As defined in Section~\ref{sec:model}, the exploration process is modeled as dynamic interplay between designs and subspaces that structure the navigation of the designs.
\prompthistwo uses a tree-like structure to visualize the relationships between different designs and subspaces.
Each node represents a design (prompt) or a subspace, and the links indicate the child node is a refinement of or inspired by the parent node.
The layout of the nodes is manually controlled by the user to maintain the user's mental model about the exploration process.

\begin{figure}[ht]
\centering
\includegraphics[width=.75\linewidth]{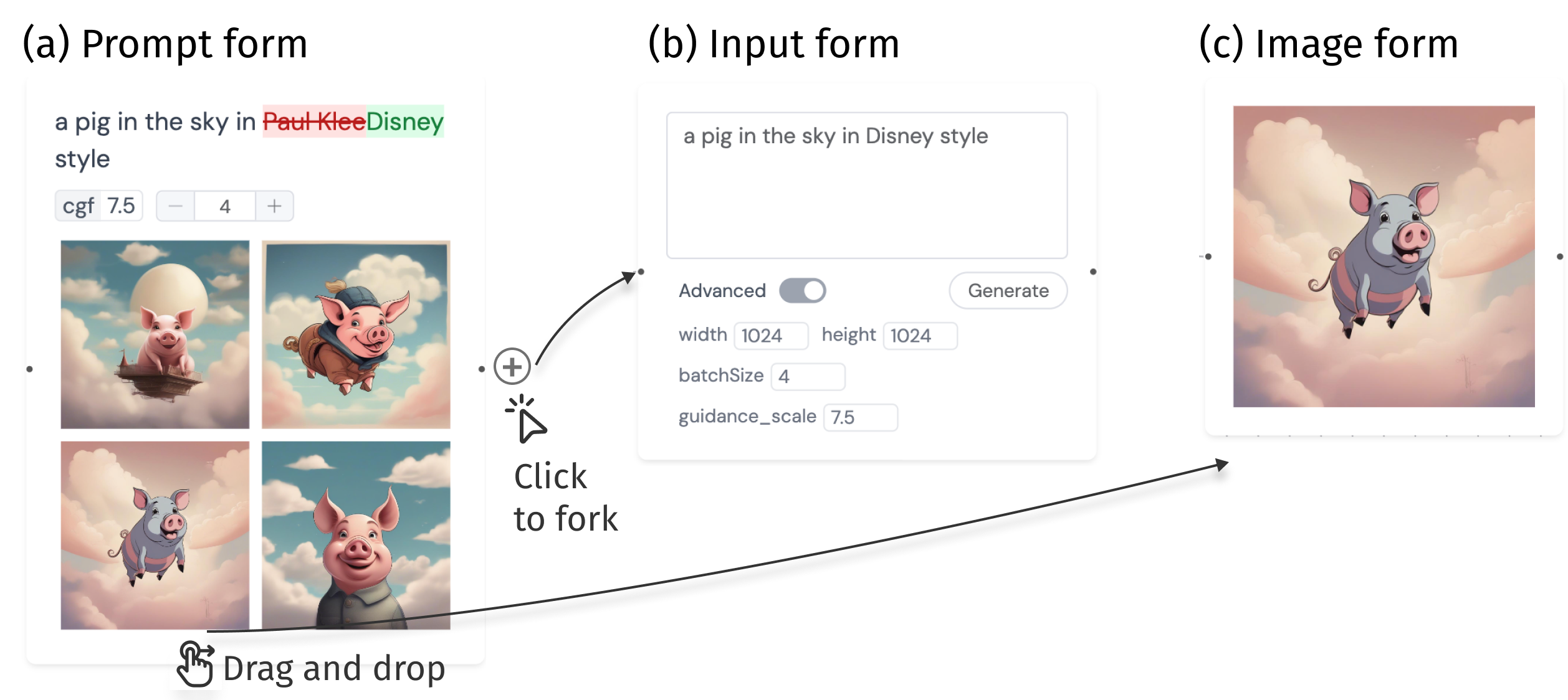}
\caption{
    Three forms of the prompt node. (a) Prompt form, which includes the prompt, parameters, and generated images. (b) Input form for changing the prompt and parameters. This is used to fork an existing node with duplicated prompt and settings to start with. (c) Image form containing a single image that the user is particularly interested in, which is created by dragging an image out of the prompt node (or the subspace node) for further exploration.
}
\label{fig:prompt-node}
\end{figure}

The \textbf{prompt node} has three forms of representation. As shown in Fig.~\ref{fig:prompt-node}, the \textit{prompt form} represents one attempt. It includes the prompt text, parameter settings (such as the number of generated images), and the output images. The changes in the prompt compared to the parent (if it exists) are highlighted. The \textit{input form} allows the user to set the prompt and parameters for the generation. It can be created either as a root node or as a child of an existing node. For the latter case, the prompt and settings of the parent node are copied to the input, so that the user can easily edit the prompt. Each prompt often generates multiple images and they are not always equally relevant to the task. The \textit{image form} contains the particular image that the user is interested in and can be created by dragging the image out of a prompt or subspace node.

\begin{figure}[h]
\centering
\includegraphics[width=\linewidth]{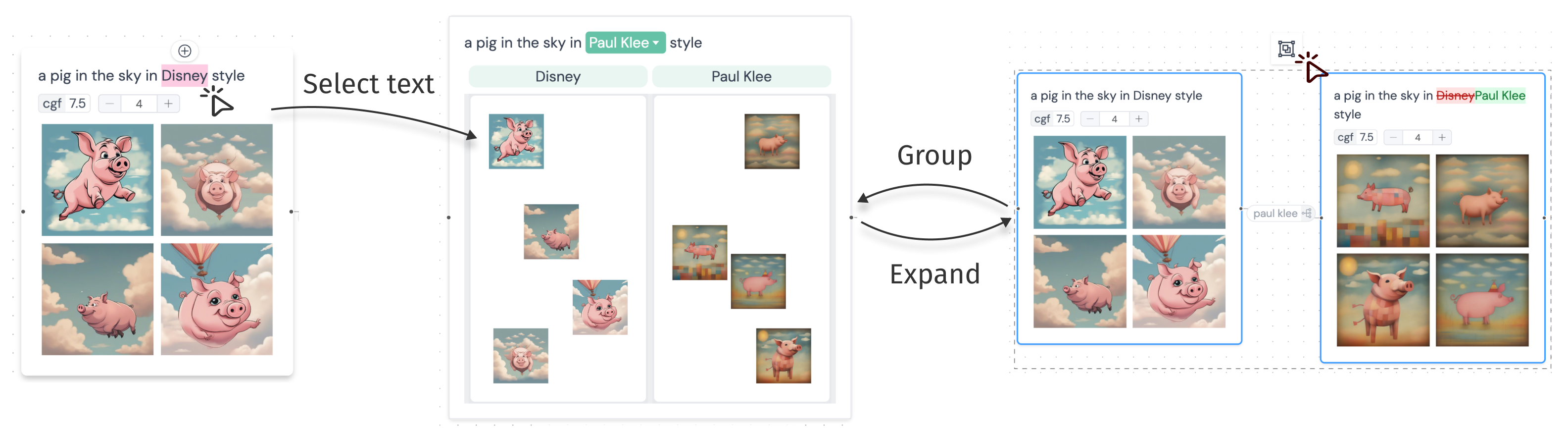}
\caption{
The subspace node can be created by adding dimension to a prompt. This is achieved by selecting some text and setting it as a dimension in the context menu. Multiple dimensions can be added to a prompt. Subspace node is a more compact representation of a collection of related prompt nodes: each addition of a dimension or its value can be alternatively represented as a child prompt node. Subspace node supports the interaction of expanding into a series of corresponding prompt nodes.
}
\label{fig:construct-subspace}
\end{figure}

The \textbf{subspace node} represents the structured subspace defined in the \demodel. We provided two ways to \textbf{construct a subspace} (Fig.~\ref{fig:construct-subspace}), i.e., by defining dimensions in the prompt text or by grouping a number of prompt nodes if the changes can be described as a dimension value change. Conversely, it can be expanded to a series of prompts nodes with each node representing one change in one dimension. The prompts in the subspace node are automatically generated, using the initial prompt as the template and replacing the original values according to the list of values for each dimension. In the current system, the dimensions are manually specified by the users. We will describe the encoding in more detail in Section~\ref{sec:grid}.

\begin{figure}[ht]
\centering
\includegraphics[width=\linewidth]{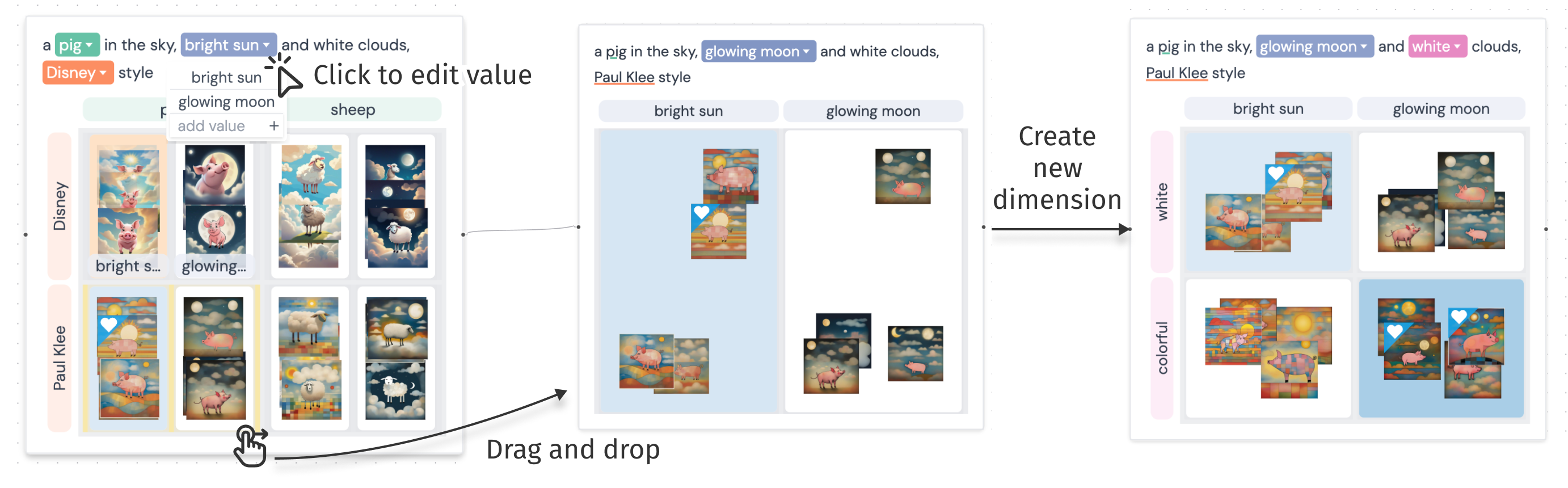}
\caption{
    The grid view applies \textit{dimensional stacking}~\cite{ward1994xmdvtool} to visualize the subspace. The space is divided recursively (nested grids) to accommodate more than two dimensions. In the top-left grid, the color green (subject) and orange (style) are the first two dimensions, mapped to the \textit{x} and \textit{y} axis respectively. The third dimension, which is  ``scene'' shown in blue, is shown by further dividing each grid cell. The cell in the grid (top left) can be dragged out as a child node (bottom left) to explore even more dimensions (bottom right). Users can either click the drop-down list embedded in the prompt or open the setting panel (top right) to edit the values.
}
\label{fig:grid}
\end{figure}

\subsubsection{Subspace Node}
\label{sec:grid}

\definecolor{dimGreen}{HTML}{66C2A4}
\definecolor{dimOrange}{HTML}{FC8E62}
\definecolor{dimPurple}{HTML}{8EA0CB}
\definecolor{dimPink}{HTML}{E78AC3}

The grid view is designed to visualize the prompts and images within the subspace. A lot of visualization methods have been developed for presenting multi-dimensional data~\cite{dzemyda2013multidimensional}. In this work, we aim to support users to compare the results generated by different values of the dimensions (\textbf{G2}). Therefore, we chose the \textit{dimensional stacking}~\cite{ward1994xmdvtool} approach, i.e., recursively selecting two dimensions to divide the space into a grid. As shown in top half of Fig.~\ref{fig:grid}, \textbf{\textcolor{dimGreen}{subject}} (\textit{pig} and \textit{sheep}) and \textbf{\textcolor{dimOrange}{style}} (\textit{Disney} and \textit{Paul Klee}) are the first two dimensions that divide the space into four cells in the left node. Each cell is further divided by the \textbf{\textcolor{dimPurple}{scene}} (\textit{bright sky} and \textit{glowing moon}) dimension as the inner grid. This way, the outputs of different values are juxtaposed and easy to compare.

The dimensional stacking method can scale to around five dimensions. This can be limiting as the artists can use long prompts with ten or more dimensions. In our design, the user can \textbf{elaborate the subspace} by editing the dimensions in the grid, or dragging out the cells of the grid as a new node and adding new dimensions there. As shown in the top half of Fig.~\ref{fig:grid}, the smaller grid on the right was a cell in the larger grid on the left. In the new node, the \textbf{\textcolor{dimGreen}{subject}} is \textit{pig} and the \textbf{\textcolor{dimOrange}{style}} is \textit{Paul Klee} (encoded with text underline), the same as the values in the cell on the left. Besides the \textbf{\textcolor{dimPurple}{scene}} dimension, the user adds another dimension: \textbf{\textcolor{dimPink}{color}}. By dragging the cell out, the users can select the values that they are interested and further refine the structure.

\subsubsection{Encoding User Curation}
\label{sec:encode-curation}

\begin{figure}[ht]
\centering
\includegraphics[width=.6\linewidth]{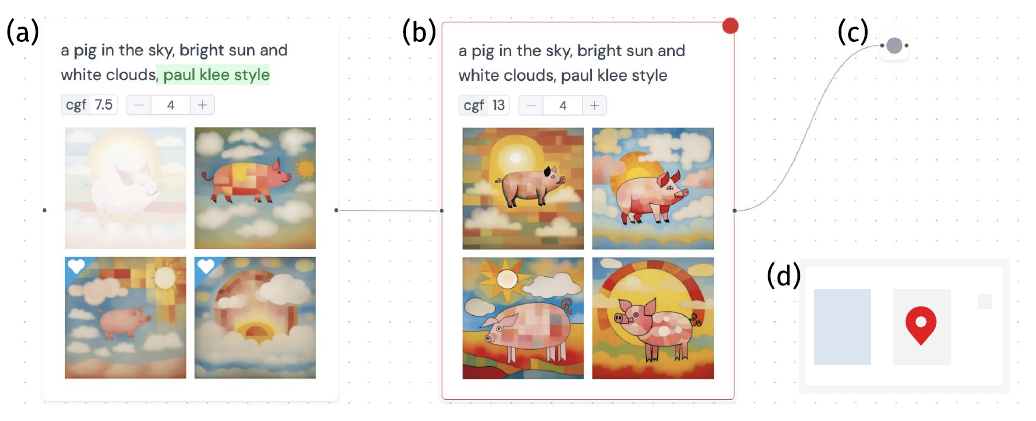}
\caption{
    User curation: (a) Like images have a heart icon and disliked images have reduced opacity; (b) Pinned node; (c) Minimized node. In the mini-map (d), the degree of like/dislike for a node is encoded with color (blue for ``like'' and orange for ``dislike'') and a location icon is added to each pinned node.
}
\label{fig:curation}
\end{figure}

Our system supports both image-level curation and node-level curation. At the image level, the user can select whether he/she likes or dislikes an image. The disliked image is encoded with lower opacity and the liked image is marked with a heart icon (Fig.~\ref{fig:curation}a). We also use color to encode the average degree of user preference on the images generated by a prompt, i.e., the color of the node on mini-map (Fig.~\ref{fig:curation}d) or the background color of the cell in the grid (Fig.~\ref{fig:grid}). Blue represents like, and orange represents dislike. At the node level, the user can pin a node if wanting to visit it later, which is encoded with a red border (Fig.~\ref{fig:curation}b) and a location mark on the mini-map (Fig.~\ref{fig:curation}d). Besides, users can minimize a node if it is irrelevant (Fig.~\ref{fig:curation}c).

\section{User Evaluation}

We conducted in-depth interviews with eight users to understand how \prompthistwo assists users in their creative exploration process. This section introduces the participants, settings, process, analyses, and results of the interviews.

\subsection{Participants and Settings}

We invited eight participants (P1-P8) for the interviews (4 female, 4 male; 4 aged between 18 and 35 years, 4 between 36 and 55 years). P1-P4 are university professors who study, teach, and practice visual arts. P1 majors in digital art, P2 is a filmmaker, P3 is a creative director, and P4 practices interaction design. P5 is an undergraduate student majoring in digital media. P6 and P7 are amateur users who are interested in visual art and have won awards in artistic competitions. P8 is a professional designer majoring in information and interaction design. All the participants frequently use generative models to explore and produce creative ideas. The models that the participants use most often include Midjourney~\cite{midjourney} (through its web interface or Discord chat) and Stable Diffusion~\cite{stable-diffusion} (through web interface or computational notebooks). The web interfaces used commonly keep the prompt history as a linear list.

The task is to use \prompthistwo to explore artistic ideas. The exploration centers around a topic or a project that the participant proposes, which mimics their usual workflow. To encourage think aloud and reduce the impact of unfamiliarity with our system, we used a collaborative evaluation approach inspired by the Pair Analytics~\cite{pair_analytics}. The system was running on the researcher's screen and the control was shared with the participants. The participants would ask the researcher to perform more complex interactions or new functions they are not familiar with, such as modifying the prompt or to create a dimension. The participants would perform simpler interactions, such as prompt input, or more complex functions when they feel comfortable to do so. This reduces potential fictions introduced by communicating every actions to the researchers.

\subsection{Process}

The interview took around one hour. It began with collecting the participant's demographic information, such as their background and experience with generative models. Then, there was a 10-min demonstration of the system to get the participants familiar with the main features. This was followed by a \textit{think-aloud} session where participants used the system to explore creative ideas for around 25 minutes. After that, the participants recounted their exploration. The last part is a semi-structured interview on the experience and feedback.

\subsection{Results}

The interview sessions were conducted through video conferences and were recorded and transcribed. We conducted a thematic analysis of the participant's feedback. Overall the participants were satisfied with the exploration and appreciated the design of the system.
The participants agreed that our system was easy to learn and use, and it is a more effective tool to manage the creative process than what they were using, as it represented the thinking process in a more intuitive manner and facilitated structured exploration.
In this section, we discuss how the participants think the system helped them with the exploration process and potential improvements.

\subsubsection{How does the visualization reflect the evolution of user thinking? (R1)}
We asked the participants about their feedback on the design of the visualization, i.e., the tree-like structure and the nested grids representing subspaces. All participants agree that the visualization is intuitive and easy to understand.

\textbf{Clear linkage between the ideas.}
The node-link representation effectively captures \textit{``where the ideas come from and how they are further developed''} (P1), thus clearly reflects the evolution of user's creative thinking. Such a representation not only helps participants review and recall their own thinking, but could potentially make their ideas accessible to others. \textit{``Even if I was somebody else, I could very much understand how they made it and what their thoughts were''} (P2).

\textbf{Nested grids allows for a scalable overview.}
The participants found the visualization provides an effective overview of the exploration process, even after attempting dozens of prompts. \textit{``It is like a snapshot of the creative process''} (P6) and the users can \textit{``see everything at once''} (P2). Compared to the list or gallery view that is common in text-to-image applications, the \textit{subspace node} in \prompthistwo, which compresses similar prompts and images into one node, is space-efficient while preserving the semantics.
While initially concerned about how to visualize a subspace with dozens of dimensions upon, P1 commented that this was no longer an issue with the ability to pull out a cell and continue to explore additional dimensions.

\subsubsection{How does \prompthistwo help users make sense of the generative models? (R2)}
Participants agreed that the tool help them understand how the prompt changes affect the model's generation. Most participants (except for P5, who were stuck due to the unsatisfactory capacity of the model) were able to refine their prompts based on the model's output to better align the generated images with their intensions.

\textbf{Juxtaposition supports effective comparison.} The subspace node uses the grid layout to visualize the dimensional structure. By arranging results according to different dimensions and their corresponding values, the grid makes it possible to directly juxtapose alternatives and identify the features or contrasts across conditions. This helped participants \textit{``easily compare the effects of different values and analyze which prompts were effective''} (P8).

\textbf{Coupling of exploration and sensemaking.} The dimensions defined in the subspace not only supports users to explore their ideas in a structured way, but also scaffolds the interpretation of the generated results. Participants could get in-situ feedback as they modify the subspace. Besides exploring different ideas, P6 and P8 used the subspace nodes to experiment with slightly different wordings to fine-tune the results.
\textit{``I can replace certain parts of the prompt with similar phrases and compare the results, which helps me understand how the prompts work.''} (P8)

\subsubsection{How do the interactions help with the exploration? (R3)}
The participants enjoyed using \prompthistwo to experiment with different ideas. The interactive visualizations provided a flexible and efficient way to carry out the exploration.

\textbf{Subspace editing encourages divergent thinking.}
Participants reported increased willingness to explore with different ideas when using \prompthistwo. The main reason is ``\textit{the system makes it very easy to create multiple dimensions and explore different values}'' (P1). Being aware of the functionality to create subspaces, the participants actively reflected on the potential variations of the existing prompts that could be explored.
Besides, P6 highlighted the emotional support provided by such interactions, \textit{``it was enjoyable to create the dimensions in that the node was turned into a grid after I added different values''}. This makes exploring divergent ideas \textit{``an interesting and playful experience''}.

\textbf{Fluid interaction reduces distractions and cognitive load.}
The participants liked the way to create dimensions, \textit{``when you make a new prompt, you can just double click on things and change it from the existing prompt''} (P2). This led to less distraction and the participants could \textit{``focus on editing the keywords''} and therefore engage in a \textit{``focused creative process''} (P3).
Participants also liked that \prompthistwo is flexible on how to experiment with multiple dimensions to fit their routines and preferences, e.g., either exploring a number of dimensions at a time or focusing on one or two dimensions and incrementally adding more dimensions. For example, P2 preferred the latter as \textit{``it is not overwhelming''}.

\subsubsection{How does the prompt history help user manage the exploration? (R3)}
All participants appreciated the system's support for \textbf{nonlinear} exploration through the intuitive representation and visualization that reflects their thinking process. In addition to the subspace modification, the participants found that the presentation of the entire exploration history is valuable for organizing the exploration.

\textbf{Easy to locate and go back to a previous attempt.}
Compared to their usual workflows, where the users often \textit{``get lost and couldn't find the right prompt to go back to''}, using \prompthistwo, \textit{``it is much easier to review the prompt history and locate the prompt that I would like to revisit''} (P3). Especially in a process that is often filled with trials and errors, users can effectively \textit{``identify ineffective prompts and restore to the more satisfactory ones and try new strategies''} (P5). This feature is enhanced by the curation support, i.e., users can \textit{``select results [they] like and reuse them or continue with the exploration later''} (P8).

\textbf{Reflection encourages divergent  and innovative thinking.}
Looking back on her exploration history, P1 found an interesting idea in the early attempts and created a branch to further develop that idea. Similar instances to \textit{``take on a different route''} (P2) were reported by the most participants. P4 suggested that the reflection support is particularly valuable for artists, \textit{``it encourages me to break out of my typical style and seek for innovation''}.

\textbf{A layer for meta-creation.}
Some participants view the organization of the prompt history itself as a kind of creation, \textit{``it added another layer of creation beyond text-to-image generation''} (P6). The canvas provides \textit{``an unlimited space''} for ideation  and allows participants to \textit{``flexibly manage a wild variety of ideas''}. The artists saw a particular value in this \textit{``documentation of the creative process''} (P1), which preserves the context of the artist's thinking and complements the final presentation of the artwork.

\subsubsection{How can \prompthistwo be improved?} We asked the participants about their feedback on the limitations of the system and how it can be improved to accommodate their workflows. A common request is the support for image-to-image generation, and they believe the value of the system would be even more significant with this feature integrated. ``\textit{I believe the tool will be even more valuable in supporting working continuously on the generated images, as it is easy to review the versions at different levels and jump to different stages}'' (P1).
Another interesting suggestion is the support for modular workflow proposed by P3. Sometimes the user creates a prompt and wants to use it as a template later. Currently this is supported by forking the node or directly modifying the dimensions. P3 suggested allowing users to save it as a template together with the possible values so it can be used in other sessions, ``\textit{so that I do not have to start from scratch, which would significantly reduce the cost of prompt engineering}''. We see this need as a way to schematize user's knowledge about text-to-image generation and preserve the knowledge for future reuse.

\section{Discussion and Future Work}

\textbf{Extending the Evaluation.}
In this work, we evaluated the work with users through qualitative interviews. The interview results provided us with in-depth understanding of how our design supports users in creative exploration. However, there are more evaluation questions that can better answered with quantitative approaches, such as 
\textit{How does \prompthistwo influence user's exploration patterns?}
The participants reported that, compared to their usual workflows, they felt more organized and motivated to think divergently when using our system for exploration. We plan to investigate this further
through comparative experiments and quantify the depth, breadth, and other metrics that characterize the exploration.

\textbf{Applying the \demodel.}
Although we mainly discussed the \demodel in the context of exploratory design and text-to-image generation, the idea of the model, as well as the design of the \prompthistwo interface, can be applied to other scenarios such as exploratory data analysis and the exploration stage of sensemaking. What may differ across different scenarios is that the structure of the subspace needs to be adapted to the task. Also, in this work, we used the same structure (subspace) for ``navigation'' and ``interpretation''. In more complex tasks (e.g., data analysis), the frame for exploration may not be sufficient for sensemaking. Further research may be conducted to investigate the interactions between different abstraction levels of the structures.

\textbf{Process visualization for human-AI interaction.}
This work focuses on the representation and visualization of the exploration process, and currently the users need to manually specify the prompts and dimensions. A lot of recent work has been exploring recommendations and automation to reduce fixation and enhance user thinking when working with generative models~\cite{suh2024luminate, almeda2024prompting}. Such support can be integrated into the \prompthistwo to further improve its effectiveness. We also hypothesize that the structured representation of user thinking captured by \prompthistwo can allow recommendation engines to better understand user's intention and preferences, leading to more productive guidance and better human-AI collaboration.

\section{Conclusion}

This work contributes a representation model and a visual interactive system for users to manage their creative exploration process.
We proposed the \demodel to formally represent the exploration process by characterizing the dynamic interactions between unstructured designs and structured subspaces.
Based on the representation, we developed the \prompthistwo system to visualize the exploration history, which involves the creation of new prompts and the construction as well as elaboration of subspaces. The system uses a tree-like node-link diagram to show the nature of the thinking process, i.e., where the ideas come from and how they are developed. We designed a readable and scalable representation to visualize the subspaces, i.e., the nested grids with cells that can be dragged out to extend the exploration. The system is evaluated with eight participants with diverse backgrounds and the participants agreed that the tool is effective in supporting organizing the thinking process and encouraging creative exploration. In future work, we aim to learn user intentions and preferences, and provide personalized and adaptive support for generative AI users.

\bibliographystyle{ACM-Reference-Format}
\bibliography{prompthis}

\pagebreak
\appendix

\section{Evaluation Results - Participants' Activities}

Here we present the temporal distribution of the participants' activities when they used \prompthistwo during the interviews, along with a preliminary analysis of the styles and patterns of their exploration.
We collected the session data created by the participants, including the nodes, prompts, and images (their creation time were recorded). The creative activities, i.e., creating prompt node, creating subspace node, creating dimension, and creating value, were reconstructed using the time of the nodes and prompts.

\begin{figure}[ht]
\centering
\includegraphics[width=.75\linewidth]{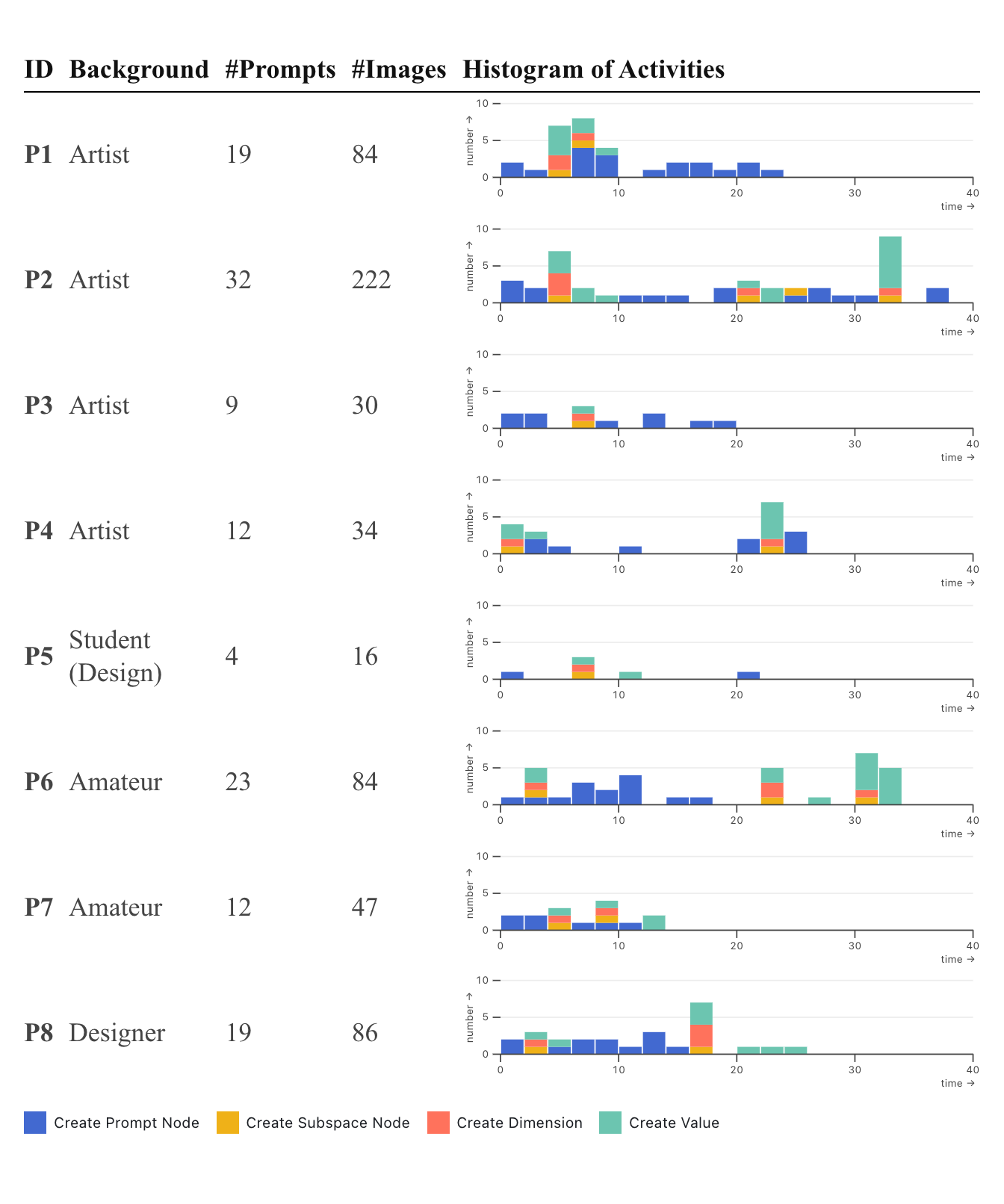}
\caption{The number of prompts and images that participants created during the interviews and the histogram of participants' activities.}
\label{fig:result-histogram}
\end{figure}

Fig.~\ref{fig:result-histogram} shows the histogram of the creative activities for all participants. P1, P2, P6, and P8 were most engaged in the creative experience.
P3, P4, and P7 had a fair level of engagement. They stopped the exploration as they thought it was beyond the model's capability to further improve the results. P5 liked the interface but found it difficult to use the base SDXL model that the system is connected to, as she usually uses LoRAs (fine-tuned Stable Diffusion models) in the daily workflow.
We observed two exploration patterns from the temporal activities of the participants:

\begin{itemize}
    \item \textbf{Divergence-convergence cycles.} The participants typically created subspace nodes and elaborated dimensions to explore different objects and styles (\textit{diverge}) after creating the initial prompts. Then, they would select satisfactory ones to further refine the prompt (\textit{converge}) by forking the prompt node. P1, P2, P4, and P6 continued to diverge to explore more ideas as the refinement was unsatisfactory or unexpected elements were observed. On average, participants created 5.89 prompt nodes between two subspace nodes (consecutive subspaces are treated as one), and the proportion of the subspace nodes among all nodes is 17.43\%.
    \item \textbf{Systematic fine-tuning.} We observed that P6 and P8 also used the subspace nodes to systematically fine-tune their prompts. After reaching useful results, they created dimensions to experiment with synonyms of certain keywords and used the grid visualization to compare the outcomes. This can be seen as \textit{elaborating the subspace} in our \demodel.
\end{itemize}

\end{document}